\documentclass[%
preprint,
 amsmath,amssymb,
 aps,
prb,
floatfix,
]{revtex4-1}

\usepackage{graphicx}
\usepackage{dcolumn}
\usepackage{bm}
\usepackage{hyperref}

\DeclareMathOperator{\erf}{erf}

\begin{document}


\title{Probing the validity of the diffuse mismatch model for phonons using atomistic simulations}

\author{Rohit R. Kakodkar}
\author{Joseph P. Feser}%
 \email{jpfeser@udel.edu}
\affiliation{%
Department of Mechanical Engineering, University of Delaware, Newark, DE, 19716 USA
}%

\date{\today}

\begin{abstract}

Due to it's simplicity the diffuse mismatch model (DMM) remains a popular description of phonon transmission across solid-solid boundaries.  However, it remains unclear in which situations the DMM should be expected to be a valid model of the underlying physics. Here, its validity is investigated mode-by-mode using a 3-dimensional extension of the frequency domain, perfectly matched layer (FD-PML) method, to study the interface between face-centered cubic solids with  interdiffused atoms. While submonolayer levels of interdiffusion are found to increase the number of available modes for transmission, consistent qualitatively with the DMM, we do not find quantitative or qualitative convergence toward the DMM at higher levels of interdiffusion.  In particular, contrary to the fundamental assumption of the DMM, modes are not found to lose memory of their initial polarization and wavevector.  The transmission coefficients of randomly interdiffused and smoothly-graded interfaces are also compared.  While smoothly graded interfaces show strong anti-reflection properties, selection rules still prohibit transmission of many modes, whereas interdiffused interfaces are not subject to such rules and achieve similar thermal interface conductance by transmitting with lower probability but using a wider range of modes.  

\end{abstract}

\pacs{Valid PACS appear here}
\maketitle



\section{\label{sec:Introduction}Introduction}

At the interface between two atomically conformal solids, discontinuities in the vibrational properties result in a finite thermal boundary conductance (TBC) for phononic heat transport.  The TBC becomes especially important whenever multiple interfaces are closely spaced together or when heat originates from sources which are highly confined by interfaces.  Examples of the former include superlattices,  thin films, and high density nanocomposites.  Examples of the latter include finFET transistors, phase-change memory, and the plasmonic transducer of a heat-assisted magnetic recording head \cite{JAP_2003_93_2_793}. Despite the technological importance of engineering TBC, there are still no widely accepted theories for predicting TBC $a \ priori$ that are in excellent agreement with experiment, though modern theories based on the diffuse mismatch model (DMM) do generally predict the correct order of magnitude over the entire range of temperature.  There are generally two well-developed theoretical frameworks that have been established for modelling TBC: (1) direct non-equilibrium computational simulations in the time-domain by molecular dynamics and (2) phonon-gas treatments which attempt to determine the transport contribution from each phonon mode.  Landry et al \cite{PRB_2009_80_16_165304} have compared TBC predictions obtained using Molecular Dynamics in the harmonic limit with those obtained using both equilibrium and non-equilibrium phonon gas approaches using transmission coefficients ($\tau$) determined through a Lattice Dynamics based scattering boundary method. Their results only show good agreement between the near-equilibrium phonon gas approach and the MD approach at mass ratios, $m_r>2$, while for smaller mass ratio’s they show that the non-equilibrium correction can be applied to the phonon gas treatment.  However, in conventional MD approaches, mode-by-mode information is not obtained, which limits the theoretical insights obtainable by such models. However, recent developments combining modal analysis with MD may reverse this trend\cite{NJP_17_10_103002, PhysRevB.90.134312, SciRep.23139.6.2016, JAP_119_1_015101}.   One potential advantage of the MD approach is the natural inclusion of inelastic phonon interactions whenever large displacements are present.  This manuscript focuses on the latter approach.

Using the phonon gas approach, the thermal interface conductance can be calculated by performing integration over k-space in the first Brillouin zone as \cite{APL_2005_87_21_211908}

\begin{equation}
    G=\frac{1}{(2\pi)^3}\sum_{p}\iiint_{\{v_g.\hat{n}>0\}\in BZ}\hbar\omega(v_g.\hat{n})\tau_{1\rightarrow2}\frac{\partial{f}}{\partial{T}}d^3\mathbf{k}.
    \label{eqn:Thermal conductance}
\end{equation}

Here, $\omega$ is the phonon frequency, $v_g$  is the group velocity, $\hat{n}$  is the unit normal to the interface plane, $\tau_{1\rightarrow2}$  is the energy transmission coefficient, and $f$  is the Bose-Einstein statistical distribution.  Due to the advent of density functional theory, obtaining the phonon dispersions and group velocities on either side of the interface is typically trivial, but methods of modelling transmission coefficients at solid-solid interfaces remain controversial.

Initial attempts to theoretically describe transmission coefficient included the Acoustic Mismatch Model (AMM)\cite{AMM_Kapitza}, which treats incident phonon reflection and transmission using continuum mechanics approximations.  Simple versions of this theory predict that energy transmission coefficients for phonons depend only on acoustic impedance mismatch.  However, at room temperature thermal interface conductance in most materials is dominated by phonons with wavelength comparable to interatomic distances.  Thus, the AMM is generally not appropriate for room temperature solids, though it may correctly describe the behaviour of some low frequency phonons that are significant at low temperatures.  Lattice Dynamics (LD) models\cite{PRB_40_6_3685,JAP_2005_97_2_024903} and the Scattering Mediated Acoustic Mismatch Model(SMAMM)\cite{JHT_2000_123_1_105} were therefore developed to give a more accurate treatment of phonon dispersion and transmission coefficients throughout the first Brillouin zone, including the short wavelength phonons absent in the AMM.  However, these models generally still employ an idealized model of the interface: smooth and epitaxial.  Fundamentally, these approaches assume coherent transport of phonons across the interface and do not account for inelastic phenomena.  

The diffuse mismatch model (DMM) was developed to handle the opposite extreme \cite{APL_1987_51_26_2200}. In the DMM, it is assumed that the interface is disordered such that, whatever the details of the interfacial structure and bonding, incident phonons are scattered so intensely that they lose memory of the polarization and direction from which they originated. Using the principle of detailed balance and conservation of energy, the transmission coefficient of an incident phonon depends only on its incident energy and is given by

\begin{equation}
    \tau_{1\rightarrow2}(\omega)=\frac{I_1(\omega)}{I_1(\omega)+I_2(\omega)},
    \label{eqn:DMM_transmission}
\end{equation}

where

\begin{equation}
    I_j(\xi)=\sum_p\iint_{\mathbf{k}:\omega(\mathbf{k})=\xi}\frac{(v_g^j(\mathbf{k},p).\hat{n})}{||v_g^j(\mathbf{k},p)||}d\mathbf{S}_{\mathbf{k},p}.
    \label{eqn:Transmission_Integral}
\end{equation}

A numerically discretized version of this equation was given by Reddy \cite{APL_2005_87_21_211908}, previously.  The integration in Eq. \ref{eqn:Transmission_Integral} is to be performed over k-space surfaces with constant frequency and polarization.  For isotropic phonon dispersions in the low frequency limit, where $\omega=c_p||\bf{k}||$  , the isofrequency surfaces are spheres, and the simpler expressions originally given by Swartz and Pohl are recovered \cite{APL_1987_51_26_2200}.   An important prediction of the DMM according to Eq. \ref{eqn:DMM_transmission} is that the phonon transmission probability is purely a function of frequency.

The diffuse mismatch model does not consider the details of interface structure.  However, a variety of computational approaches have been developed that can simulate phonon-structure interactions considering specific atomic placements/bonding at the interface.  The two most highly developed of these are Molecular Dynamics (MD) simulations and the Atomistic Green's Function method (AGF) \cite{PRB_2003_68_23_245406}.  These methods have previously been used to study a number of microscopic effects that are important to thermal boundary resistance, including pressure\cite{PRB_2011_84_19_193301,PRB_2011_84_19_195432}, interfacial bond strength\cite{PRB_2009_79_10_104305}, grain boundaries\cite{JAP_2004_95_11_6082}, and lattice mismatch between the materials interfaced \cite{Stevens20073977}.  Molecular dynamics can be used either to directly simulate macroscopic transport properties\cite{PRB_2000_61_4_2651} or to test the mode resolved transmission properties of energy across an interface.  In the latter case, wavepackets are prepared and their interaction with a structure is tracked in the time-domain \cite{APL_2002_80_14_2484,PRB_2008_77_9_094302}.  Wavepacket MD can present some challenges though, since a wavepacket necessarily contains a spectrum of phonon frequencies. To accommodate a wavepacket, the simulation domain size must be larger than the wavelengths to be simulated, and the wave must be tracked long enough to complete its interaction with the structure, which can be a challenging for highly dispersive modes.  The atomistic Green's function method has no such restrictions, but considers only harmonic bonding and is fairly limited in terms of its maximum domain size.

The approach here is to utilize an atomistic, mode-resolved computational method to directly simulate the phonon transmission coefficient of the interface between face-centered cubic solids with interdiffused atoms.  The purpose is to directly test the validity of the diffuse mismatch model mode-by-mode, and to characterize the transition from specular to diffuse transmission behaviour as a function of disorder parameters, with the goal of trying to understand when an epitaxial interface should be considered smooth or rough from the the point-of-view of phonon transport.  Previously, several other works\cite{PRB_2012_86_23_235304, Shao201533, C5NR06855J} have found that small interdiffusion layers result in higher thermal interface conductance. Tian et al. \cite{PRB_2012_86_23_235304} attributed this enhancement in TBC to increased density of states overlap due to interdiffusion. In accordance with this hypothesis, English et al. \cite{PhysRevB.85.035438} found that thermal boundary conductance can also be increased by inclusion of an intermediate layer with mediating vibrational properties (achieved by varying the mass of the intermediate layer). Furthermore, Zhao et al \cite{JAP_2009_105_1_013515} studied geometrically roughened Si/Ge interfaces between FCC crystal structures. Their results indicate that the overall thermal conductance is independent of roughness parameters. 

While several atomistic computation methods are available, we use the recently developed frequency-domain perfectly matched layer (FD-PML) method \cite{JAP_2015_118_9_094301}, rather than the Atomistic Green's Function (AGF) approach or wavepacket MD, since the FD-PML technique naturally concerns the energy reflection/transmission coefficient of a single incident wavevector and polarization. 



\section{\label{sec:Methodology}Methodology}

The FD-PML method~\cite{JAP_2015_118_9_094301} involves frequency domain decomposition of the atomistic equation of motions of each atom, and solving for a scattering system’s response to an incident plane wave of specified wavevector and polarization.  A perfectly matched layer (PML) is used on the boundary of the scattering system to damp scattered waves without spurious reflections of the scattered waves back into the scattering system. Thus, a well-designed PML makes the boundary of the scattering system indistinguishable from a boundary that allows infinite propagation. Like the AGF method, FD-PML reduces scattering problems to a system of sparse, banded linear algebraic equations, but FD-PML has better scalability to large system sizes since it does not require storage of a fully inverted matrix and can thus utilize efficient, low memory iterative solver algorithms.  In our original description of the FD-PML method, the method was demonstrated in one- and two-dimensions~\cite{JAP_2015_118_9_094301}.  Details of its generalization to 3D for arbitrary crystal structures and bonding are given in an Appendix.
\begin{figure}
\begin{center}
\includegraphics[width=0.33\textwidth]{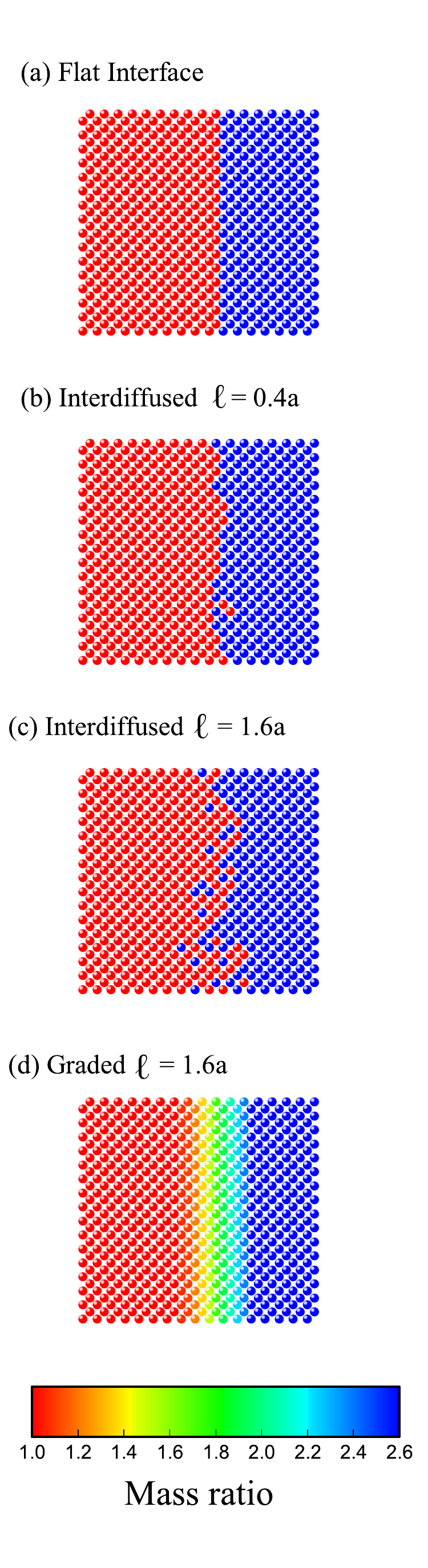}
\end{center}
\caption{\label{fig:Mass_graphs} Cross-section of the simulated interface. The colorbar indicates the mass ratio $m/m_{Si}$}
\end{figure}
 Disorder is introduced using random interdiffusion of the atomic masses from each interface (Fig.~\ref{fig:Mass_graphs}).  To emulate physical interdiffusion, the masses of each atom are assigned with probability,
 \begin{equation}
     P_1(z)=\frac{1}{2}\left[1-\erf\left(\frac{z-z_0}{\ell\sqrt{2}}\right)\right],
 \end{equation}
 where $z_0$ is the location of ideal interface and where $P_1$ is the probability of finding an atom of type $1$ at location $z$; $\ell$  is the characteristic interdiffusion length and $\erf$ is the error function.  By definition, if an atom type is not 1, then it must be 2.   To maintain simplicity, we study interfaces between FCC crystals.  Our base case is an idealization of a Si/Ge interface.  True Si/Ge takes the diamond crystal structure, but we adopt the approach of Tamura et al\cite{PRB_1999_60_4_2627} of modelling the two atom per primitive unit cell diamond structure as a single atom per primitive unit cell FCC crystal, by lumping the mass of two atoms into one (i.e. in this model, $m_{Si}$ = 56.2 amu, $m_{Ge}$ = 145.2 amu), treating only the nearest neighbour bonding (K=42.2 N/m), and using the average lattice constant between Si and Ge ($a=5.54\text{\AA}$).  This achieves a similar sound speed to Si and Ge, and a similar maximum frequency using a similar lattice constant to the true crystal. We should note that by reducing the number density of atoms by 2-fold, this has the undesirable effects of producing an artificially low heat capacity and eliminating optical phonons, which are clearly important to the real transport properties.  However, the model is merely to demonstrate the basic physics of interface transport in the presence of disorder, so its realism compared to Si/Ge system should not be taken too seriously.
 
\begin{figure}
\includegraphics[width=0.45\textwidth]{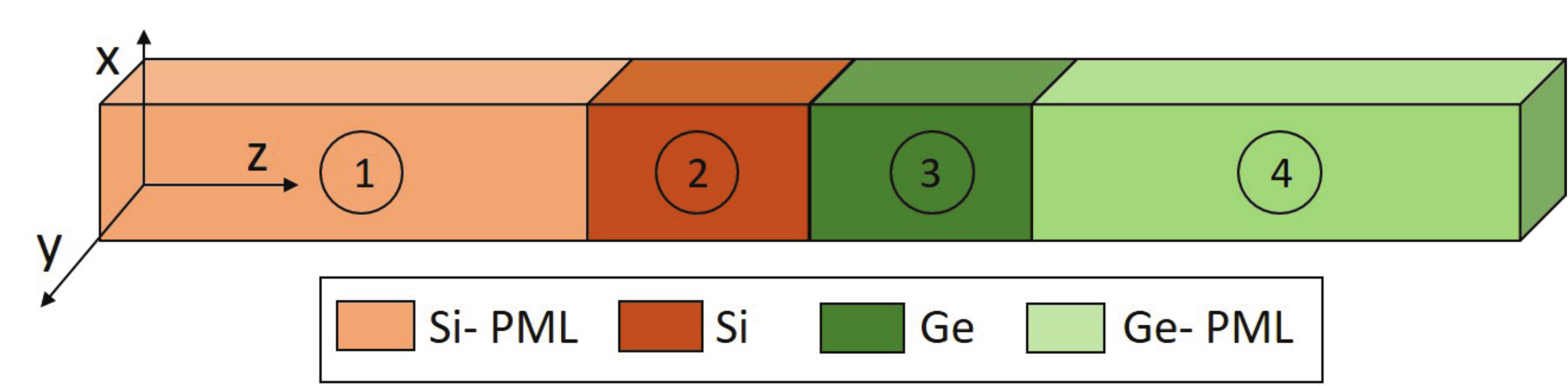}
\caption{\label{fig:Computational_Domain} Schematic of the Computational Domain. Hard wall boundary conditions are applied at edges perpendicular to z-direction and periodic boundaries are applied along all other edges.}
\end{figure}
 
 Our computational domain is a rectangular box composed of a central region containing the interface with the coordinate z running perpendicular to the interface, and bounded by perfectly matched layers of the Si and Ge in the left and right z-directions (Fig~\ref{fig:Computational_Domain}).  The z-width of the central region of the simulation box was 10 conventional unit cells (i.e. 10a).  The x-y directions of the box are taken to be periodic with a period in each direction of $L_x=L_y=16a$ .  The Born von Karman boundary conditions place restrictions on the allowable in-plane k-vectors for the simulations: $k_x=2\pi n/L_x$ $k_y=2\pi n/L_y$, $(n=0,\pm 1,... \pm L/a)$ ; there are no restrictions on the allowable $k_z$, however we perform the integration in Eq.~\ref{eqn:Thermal conductance} using a uniform mesh method spanning the upper half of the first Brillouin zone\cite{HARDY1973591}; for k-points where the z-component of incident group velocity is negative, the corresponding k-point with positive group velocity is instead simulated (i.e. $\tilde{k}\rightarrow-\tilde{k}$).  It was found that reducing the areal dimension by half of the interface did not significantly affect the calculated transmission coefficients for the range of disorder length ($\ell$) studied.  The PML region width was chosen using our previous non-dimensional optimization criteria \cite{JAP_2015_118_9_094301}, which found that larger wavelength requires larger PML domain sizes to maintain accuracy ($L\approx2\lambda$ for 2-digit accuracy of transmission coefficients).  Additional details regarding the PML design are given in the Appendix.  Thus, typical simulation boxes including the PML contained between 10,000 - 80,000 atoms, with the larger size corresponding to long-wavelength phonon simulations. For each level of interdiffusion, the energy transmission coefficients for 24,960 individual phonon modes were calculated using a uniform k-grid in the Brillouin zone.  In order to solve such a large system using minimal computing resources, we used sparse, iterative solvers. The biconjugate gradient method and the generalized minimum residual methods were both found to be effective without preconditioners.  Note that because of the roughness of the interface, there were no guaranteed symmetries in the $k_x$-$k_y$ plane, though as we will show one does nearly recover cubic symmetries for statistically large simulation domains.  Once the transmission coefficient, frequency, and group velocity of each mode are known, the thermal interface conductance can be easily calculated using Eq.~\ref{eqn:Thermal conductance}.  For the case of a flat interface, the Lattice Dynamics technique developed by Young and Maris~\cite{PRB_40_6_3685} was used to validate our transmission coefficients in a mode-by-mode manner.  That technique is specifically designed to study the phonon transmission across the interface between FCC materials with a one atom basis, bonded by nearest neighbor forces, which exactly conforms to our computational case.  For those k-points where the lattice dynamics approach is numerically stable, the predicted transmission coefficients were identical to better than 0.01.  Interface conductance for a flat interface obtained for a similar test case was reported by Zhao~\cite{JAP_2009_105_1_013515} as G=111 MW/m$^2$-K, with the only difference being the use of a bonding strength in the FCC-Ge material, 0.88 times the value used here and by Tamura\cite{PRB_1999_60_4_2627}.  We find a value of G=123 MW/m$^2$-K in this case, consistent with the lower degree of mismatch between materials.


\section{\label{sec:Results and Discussions} Results and Discussions}

\begin{figure}
\begin{center}
\includegraphics[width=0.5\textwidth]{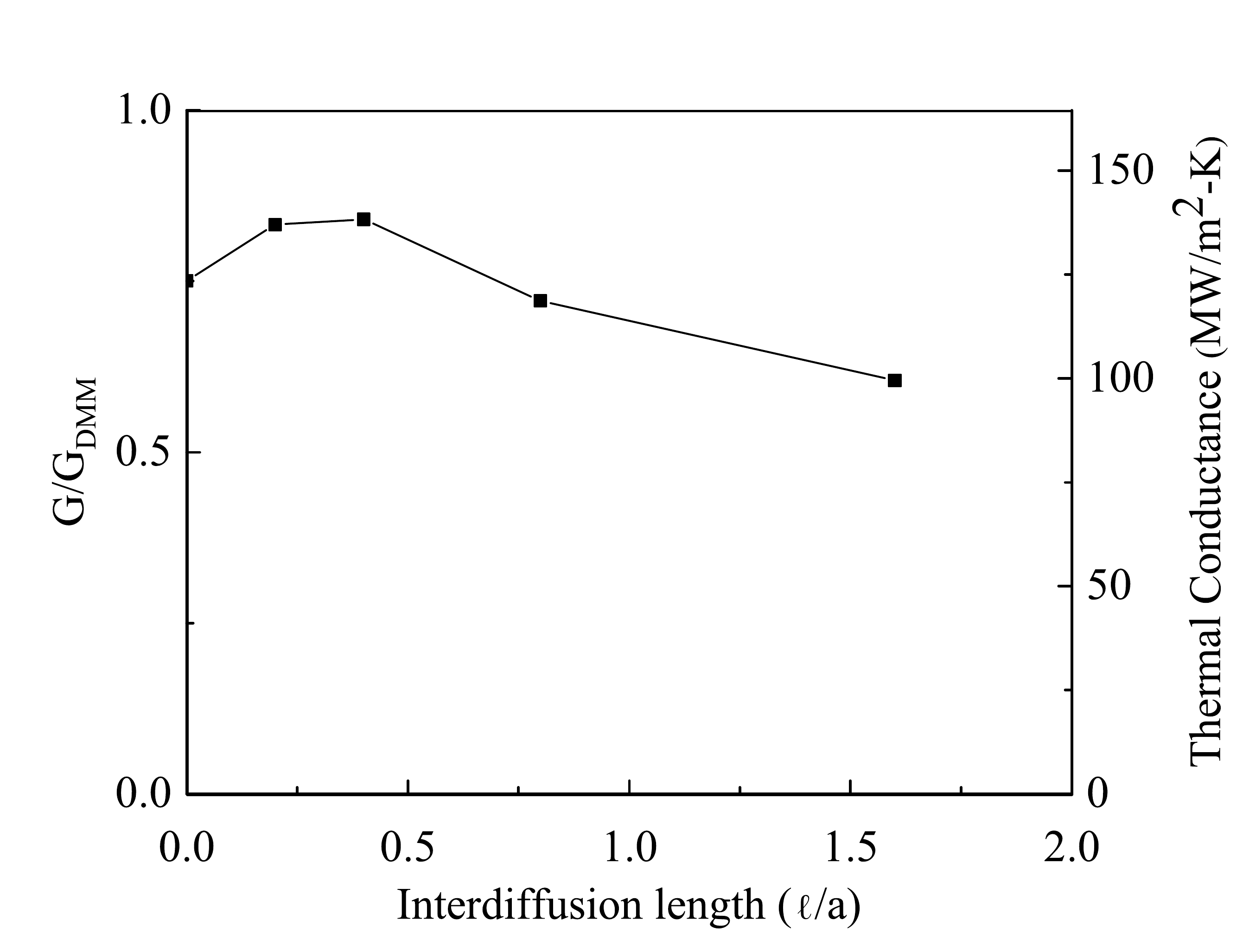}
\end{center}
\caption{\label{fig:Thermal_conductance} Normalized thermal interface conductance as a function of normalized roughness length. Here $G$ is the thermal interface conductance obtained from simulations and $G_{DMM}$ is the TBC obtained using DMM, $a$ is the lattice parameter of FCC-Si.}
\end{figure}

It is instructive to observe the collective behaviour of thermal boundary conductance prior to discussing a mode-by-mode analysis.  Figure~\ref{fig:Thermal_conductance} shows the thermal boundary conductance as a function of the interdiffusion distance, $\ell$ .  If the Diffuse mismatch model (DMM) is a valid description of transport for a disordered interface, then one expects that as disorder is increased, $G$  should tend toward the DMM value. Interestingly, Fig \ref{fig:Thermal_conductance} indicates that this is not the case.

In the large interdiffusion limit ($\ell=1.6a$), we see that the thermal interface conductance is not approaching the DMM value.  For the model Si/Ge interface here, the DMM result for $G$  is greater than that for a perfect interface.  In Fig \ref{fig:Thermal_conductance}, it can be seen that small levels of interdiffusion do increase G to 85\% of the DMM value.  However, the maximum corresponds to submonolayer interdiffusion distances, $\ell$.  Beyond this level of interdiffusion, the thermal interface conductance is observed to fall monotonically and thus does not converge toward the DMM.  In the thick interface limit ($\ell\rightarrow\infty$), one might expect $G\sim \ell^{-1}$ (i.e. rather than an interface conductance, one might think of the region as having a bulk thermal conductivity describing the resistance, $\ell/k\sim1/G$).  However, Fig.~\ref{fig:Thermal_conductance} doesn’t show convergence toward that diffusive limit either; this is clear because the rate of decrease in G is weakening with increased $\ell$.  Previously Tian et al.\cite{PRB_2012_86_23_235304} observed qualitatively similar results using AGF method, where interdiffusion at the interface of a Si/Ge (diamond structure) interface initially increased TBC followed by a monotonic decrease with greater levels of intediffusion.  However, they did not compare their results to the DMM.  In contrast, Zhou et al. \cite{PRB_2013_87_9_094303} used NEMD to characterize TBC of roughness in the form of geometrically wavy interfaces (not interdiffused), and found that $G$ monotonically increases with increasing roughness length; this was attributed to an effective increase in surface area of the interface.

\begin{figure}[h]
\begin{center}
\includegraphics[width=0.53\textwidth]{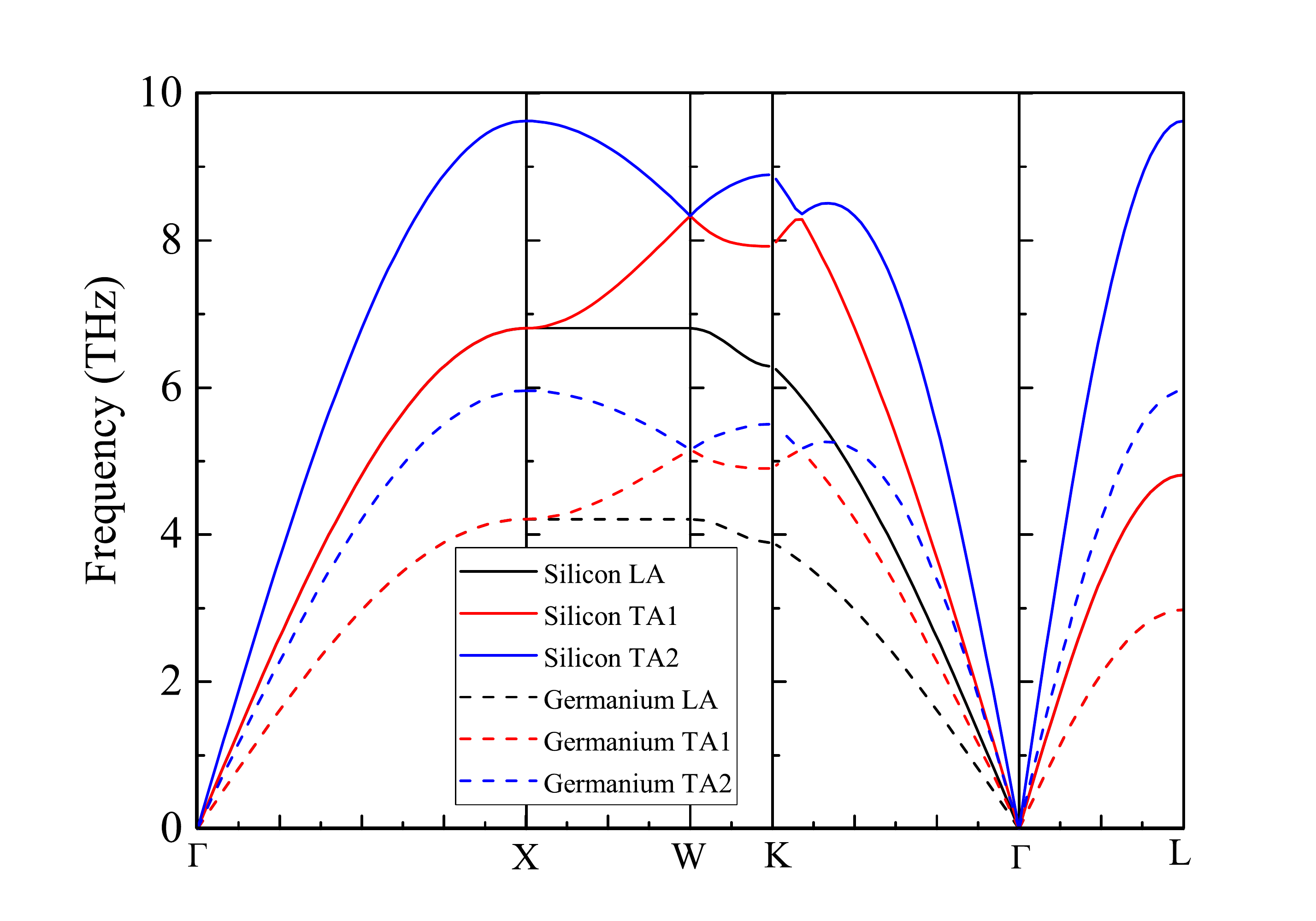}
\end{center}
\caption{\label{fig:Dispersion_relation} Dispersion relation for FCC-Si and FCC-Ge along high symmetry lines}
\end{figure}

\begin{figure}[!tbp]
    \includegraphics[width=0.53\textwidth]{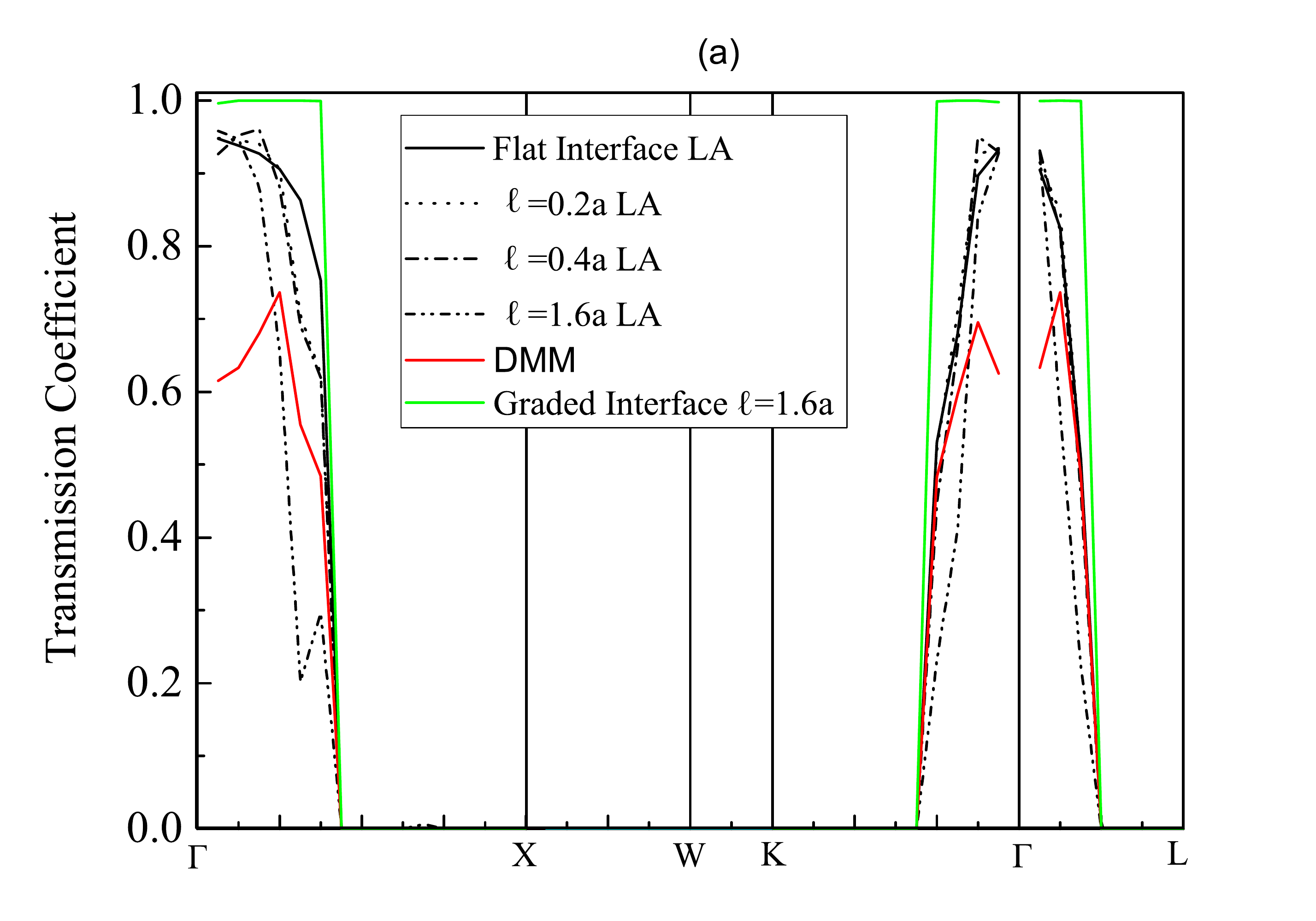}
    \includegraphics[width=0.53\textwidth]{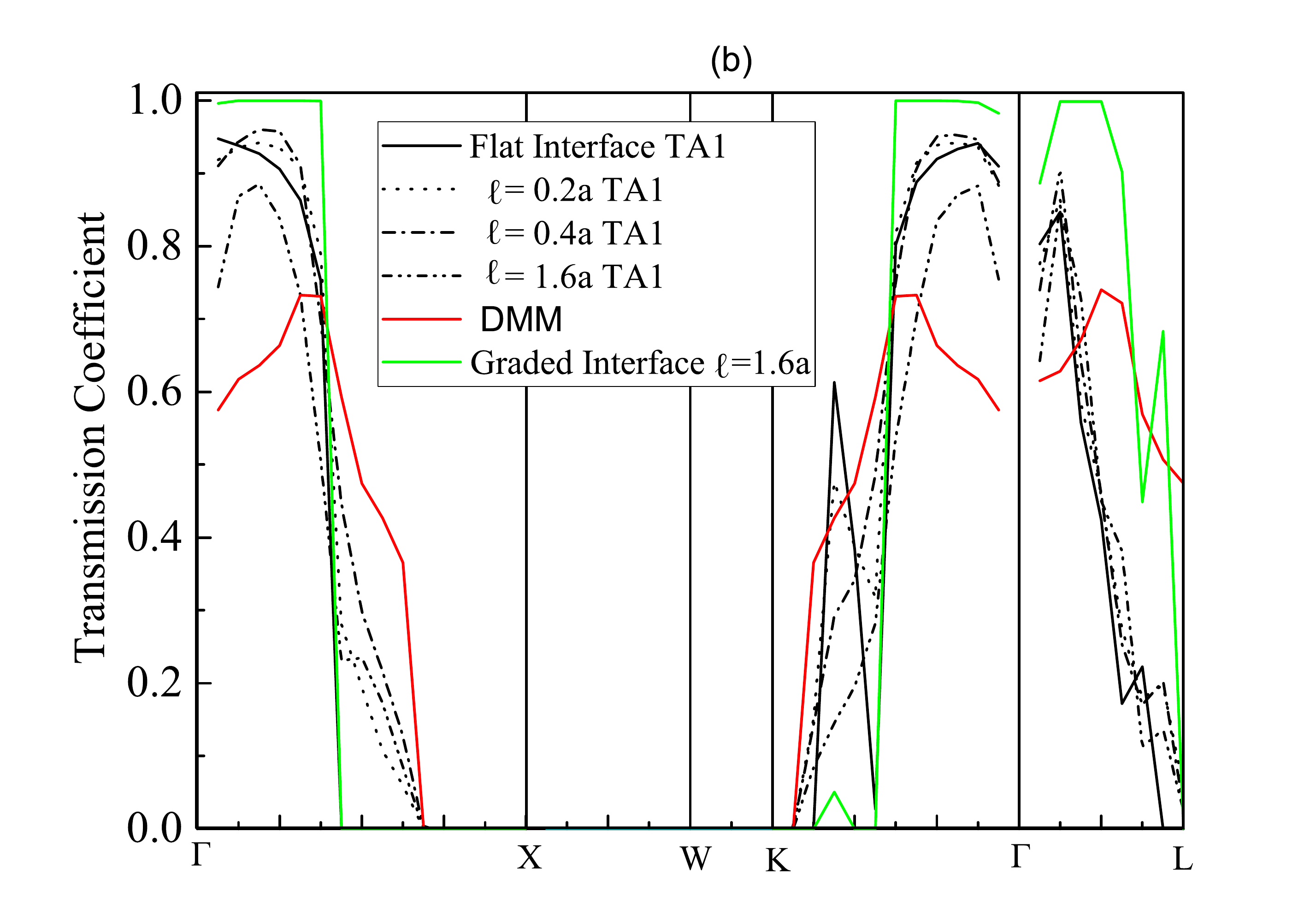}
    \includegraphics[width=0.53\textwidth]{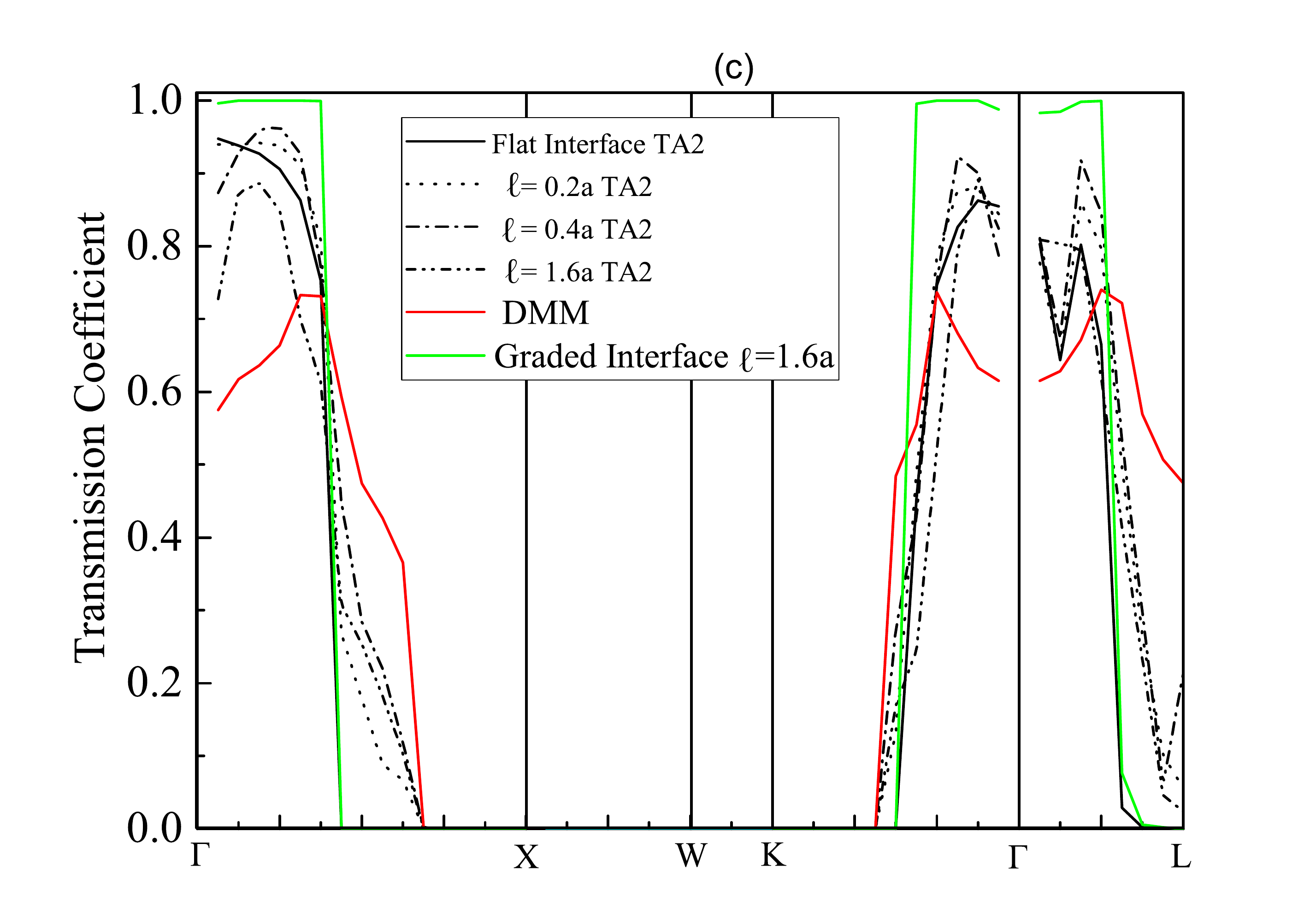}
\caption{Transmission coefficient along some high symmetry lines for the longitudinal (a) and transverse (b \& c) phonon polarizations.  For each polarization, the simulation results are given for an ideal interface, for three levels of interdiffusion and for mass-graded interface.  The prediction of the diffuse mismatch model is given for reference.}
\label{fig:High_symmetry}
\end{figure}

In order to understand the physics behind this behaviour, we analysed mode resolved transmission coefficient data over the entire Brillouin Zone.  The phonon dispersion for each material is given in Fig~\ref{fig:Dispersion_relation} along the high symmetry lines.  The corresponding phonon transmission coefficients are given in Fig~\ref{fig:High_symmetry} for phonons incident from side 1 (Si), corresponding to the longitudinal and the two transverse phonon polarizations.  The predictions for a perfectly smooth interface and the DMM are provided for comparison.  Comparing Fig~\ref{fig:Dispersion_relation} \&~\ref{fig:High_symmetry}, several important effects can be seen.

One of the most distinctive effects of disorder is to make certain types of mode conversion possible, and consequently to increase the number of available pathways for transmission.  In Figs.~\ref{fig:High_symmetry}b and 4c, an example of this can be seen in the transverse modes traveling along $\Gamma$-X:  these have identically zero transmission coefficient for a perfectly smooth interface once the frequency is higher than the maximum frequency of TA phonons on the Ge side.  However, submonolayer levels of disorder change this, enabling the high frequency/wavevector TA phonons in Si to scatter into longitudinal modes of the Ge.  However, the value of the transmission coefficients in this region never reaches as high as predicted by the DMM, and eventually decreases at higher interdiffusion levels.  The DMM, thus contains some of the right physics (i.e. increased mode conversion), but does not quantitatively match the transmission coefficient predicted by an atomistic approach.  Previously, Tian et al have attributed increase in TBC to increased overlap of density of states of the materials across the interface due to intermixing\cite{PRB_2012_86_23_235304}.  However, figure~\ref{fig:High_symmetry} indicates that this is more likely to be due an increase in the number of transmission pathways.

\begin{figure*}[p]
\begin{center}
\includegraphics[width=\textwidth]{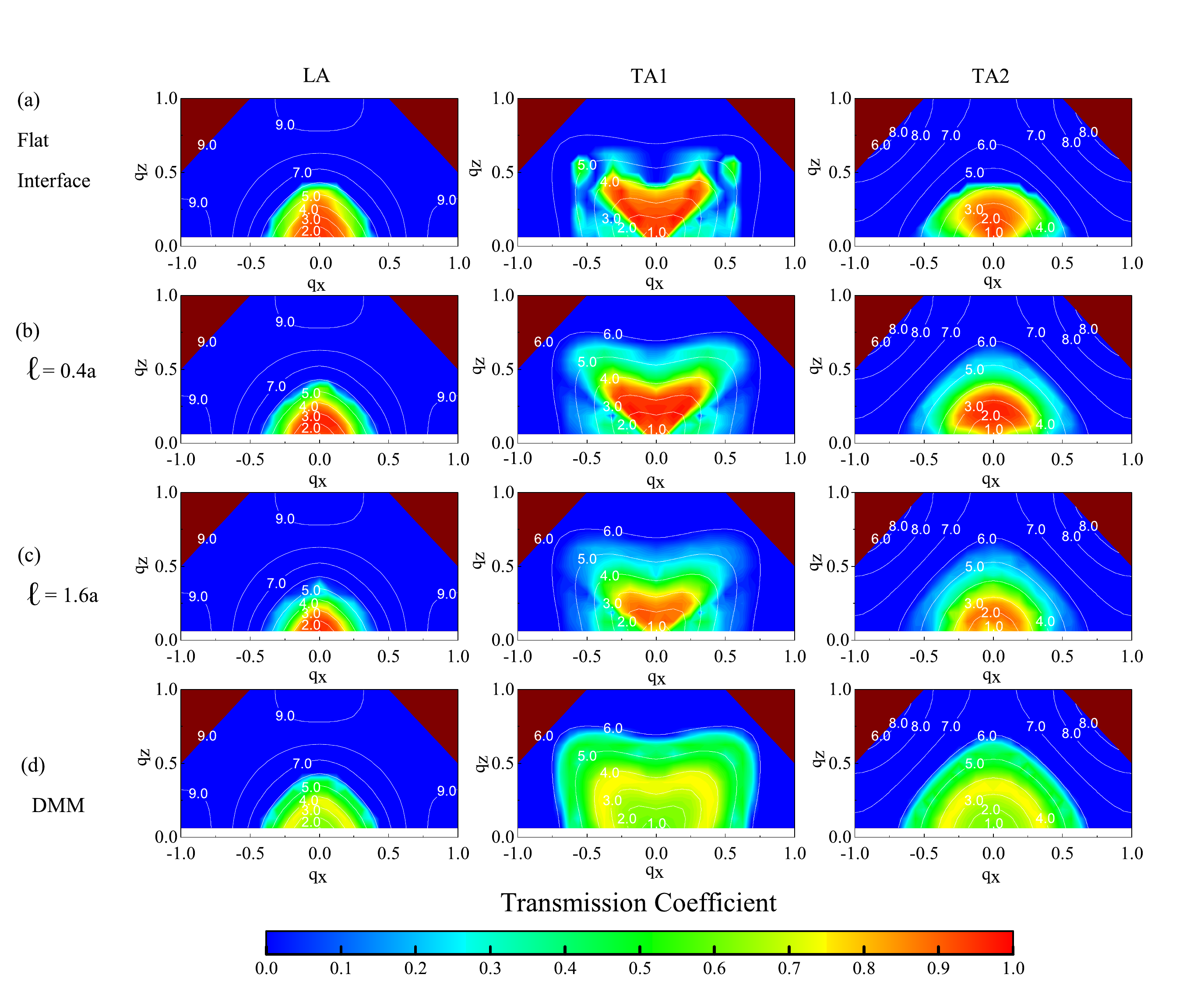}
\end{center}
\caption{\label{fig:combined_contour} (contour) Variation of transmission coefficients along the x-z plane of the BZ for FCC-Ar for a. Ideal Interface b. Rough interfaces with $\ell$= 0.4a  and c. $\ell$ = 1.6a d. DMM prediction. The colour contour represents the transmission coefficients, white contour lines indicate the isofrequency lines with their values in THz}
\end{figure*}

\begin{figure*}[!tbp]
\begin{center}
\includegraphics[width=\textwidth]{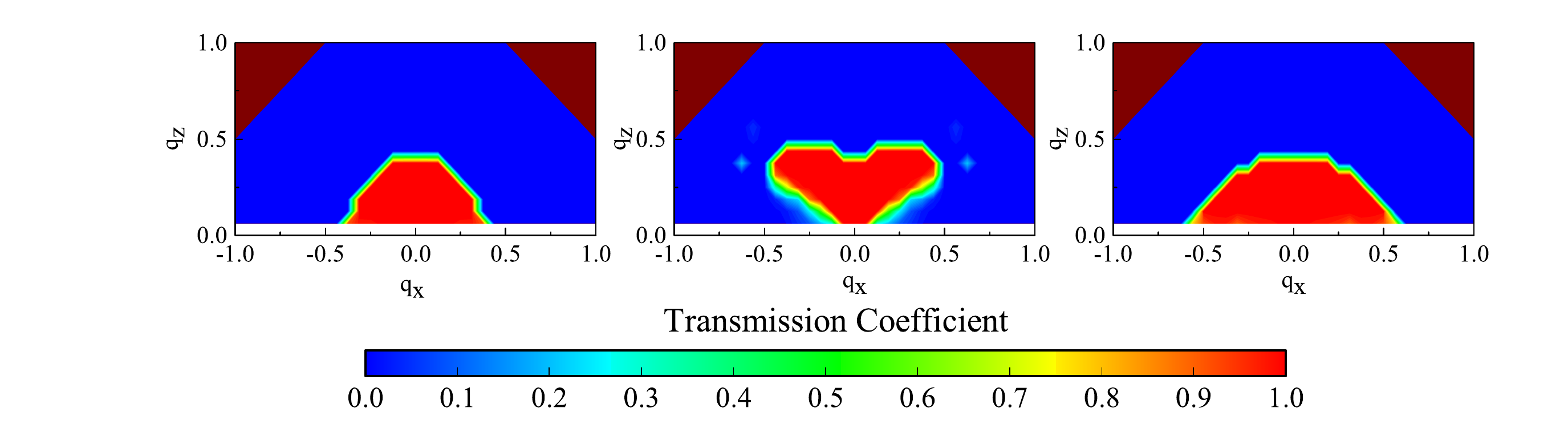}
\end{center}
\caption{\label{fig:graded_interface} Transmission coefficient for a smoothly mass graded interface with characteristic distance, $\ell=1.6a$.  For comparison, a mass gradient achieved by interdiffusion is given in Fig. \ref{fig:combined_contour}c.}
\end{figure*}

Additional violations of the DMM’s predictions can be seen throughout the Brillouin zone in Fig~\ref{fig:combined_contour}.  The DMM predicts that all incident modes with equal frequency have the same probability of transmission (see Eq~\ref{eqn:DMM_transmission}).  Fig~\ref{fig:combined_contour} compares the frequency contours in the $k_x$-$k_z$ plane of the Brillouin zone to the computed values of the transmission coefficient for three different levels of interdiffusion ( 0, 0.4a, and 1.6a top to bottom) and for all three polarizations (LA, TA1, TA2 left-to-right).  Note that the figure denotes the coordinate of the incident phonon.  For a smooth interface, the transmission coefficients do not follow the isofrequency contours, and the transmission coefficients are typically stronger for modes directed toward the interface; this is especially true for the TA modes.  The incident TA1 phonons show a peak transmission coefficient along the $\Gamma$-K direction associated with a mid-zone maximum in the dispersion relation.  Interdiffusion can be seen to have two primary effects in Fig.~\ref{fig:combined_contour}.  Fig.~\ref{fig:combined_contour}b shows that even $\ell=0.4a$ is enough to greatly broaden the participation of modes in transport.  While this low level of interdiffusion does not yield transmission coefficients that are consistent with the DMM (Fig.~\ref{fig:combined_contour}d) throughout the entire Brillouin zone, the changes are such that many of the high wavevector modes in the center of the Brillouin zone that dominate the thermal interface integral (Eq.~\ref{eqn:Thermal conductance}) do have similar transmission coefficient.  On the other hand, further increases to the interdiffusion distance (Fig ~\ref{fig:combined_contour}c) results in the general decrease in all transmission coefficients with shorter wavelength phonons being disproportionately affected.  One might attribute this to scattering, though as we've noted above the interface conductance in Fig.~\ref{fig:Thermal_conductance} does not appear to be converging to a diffusive limit, $G\sim \ell^{-1}$.  Even at the large level of interdiffusion in Fig.~\ref{fig:combined_contour}c, the transmission coefficients to not follow the isofrequency lines.  This shows quite clearly that a fundamental assumption of the DMM is incorrect at the mode-by-mode level:  a phonon mode's transmission coefficient is not independent of its incident wavevector and polarization, and dependent only on frequency (i.e. memory-less).  While one should not expect long wavelength modes to act diffusely since they do not react to local disorder as strongly, Fig.~\ref{fig:combined_contour}c shows that even many short wavelength modes maintain memory of their incident direction.

Finally, we would note that an interdiffused interface is one way to achieve a graded transition between interfaces.  Several groups have proposed that a gradation between layers could enhance thermal interface conductance either by providing a material with an intermediate matching of density of states\cite{PRB_2000_61_4_2651} (i.e. SiGe intermediate layer) or by acting as a coherent anti-reflection coating\cite{JAP_2014_116_8_083503,PRB_2015_92_14_144302}.  Due to the disorder being on the same length scale as the phonon wavelengths, the phonon transmission behavior of an interdiffused interface bares little in common with that of a smoothly graded mass transition.  To demonstrate this, we constructed a simulation where, rather than having statistically placed masses, a smooth transition in the masses (Fig.~\ref{fig:Mass_graphs}d) is assumed between materials, over a characteristic distance, $\ell$.  At any atomic site at coordinate, $z$, the mass of the atom is given deterministically by 
\begin{equation}
    m(z) = m_\textrm{Si} + \frac{m_\textrm{Ge}-  m_\textrm{Si}}{2}\left[1+\erf\left(\frac{z-z_0}{\ell\sqrt{2}}\right)\right].
\end{equation}

 This gives the same average mass profile as the the interdiffused case, but without the presence of random disorder. Fig.~\ref{fig:Mass_graphs} shows the schematic of different interfaces studied in this article. Such a gradation might be impossible to achieve practically since it would require a different homogeneous mass within each layer, but it serves as a useful point for comparison.  Figure \ref{fig:graded_interface} shows the resulting phonon transmission coefficient throughout the $k_x$-$k_z$ plane of the Brillouin zone.  It can be seen that for a continuously graded interface, the transmission coefficients are near unity for a wide range of incident phonon wavevectors. Thus, the smooth mass-gradient acts as an effective phonon anti-reflection coating for phonons with sufficiently short wavelength.    However, comparing Fig. \ref{fig:graded_interface} with Fig. \ref{fig:combined_contour}c shows that if the mass gradient is instead achieved by interdiffusion over the same characteristic distance and with the same average mass profile, the anti-reflection effect is not reproduced and the transmission coefficients are generally smaller.  Thus, these results show that a phononic anti-reflection coatings cannot be practically realised by mere interdiffusion of materials across an extended distance.  The resulting thermal interface conductance from an interdiffused interface may even be lower than what could be expected for either an ideal interface or a perfectly diffuse interface, as shown in in Fig. \ref{fig:Thermal_conductance} (see $\ell=1.6a$ ).  In addition, while a smooth mass gradient produces high transmission coefficients, the predicted thermal interface conductance (145 MW/m$^2$-K) is actually $\approx$9\% lower than the DMM (163 MW/m$^2$-K) and only $\approx$5\% higher than for the slightly interdiffused interfaces ($\ell=0.2a$ and $\ell=0.4a$ in Fig.~\ref{fig:Thermal_conductance}). This is because the smoothly graded interface with its perfect in-plane ordering appears to maintain in-plane scattering selection rules, preventing mode conversion, and effectively blocking higher wavenumber modes from participating in transport.  The interdiffused case $\ell=0.4a$ (Fig.~\ref{fig:combined_contour}c) has no such restriction and achieves similar thermal interface conductance using a wider range of transmitting modes with lower average transmission coefficient. Thus, the effect of a high transmission coefficient obtained with a graded interface does not significantly outweigh the effect of increased pathways for transmission brought about by interface disorder.

\section{\label{sec:Conclusions} Conclusions}

In summary, the validity of the diffuse mismatch model (DMM) is investigated on a mode-by-mode basis for interdiffused interfaces, using a 3-dimensional extension of the frequency domain, perfectly matched layer (FD-PML) method. It is found that disorder at an interface can increase the number of available modes for transmission, and subsequently raise the thermal interface conductance; these general observation are consistent with the DMM.  However, while submonolayer levels of interdiffusion do reproduce similar thermal interface conductance values as the DMM, the mode-by-mode predictions of transmission coefficient vary drastically from the DMM.  In particular, (1) contrary to the fundamental assumption of the DMM, not all modes appear to lose memory of their initial polarization and wavevector for the interdiffusion lengths studied here.   (2) Interdiffusion length in excess of a monolayer generally are found to make agreement between the DMM and the simulations worse, not better.  (3)  The DMM tends to overestimate the transmission coefficient of short wavelength phonons at interdiffused interfaces, and to underestimate in longwavelength regions.  

With regards to the first point, one intriguing question remains: if the modes aren’t being diffusely scattered, then what modes are they being scattered into, and how are they distributed?  At a perfect interface, there are only 3 modes of transmission possible , while in the DMM there are an infinite number, evenly distributed along isofrequency lines for each polarization.  If the DMM fails, then it must be because the scattered modes are not distributed in this way, and are perhaps prevented from doing so by additional selection rules or the physics of the scattering mechanism.  The details of how this occurs remain an open question at present, as the current FD-PML method only intrinsically tracks the incident phonon mode (wavevector, polarization), and does not give details on the modes that are involved in the scattered wave.  In principle, though, models like scattering boundary method are able to provide this information. Although it would be more natural for our method to use simple modal analysis\cite{IJHMT_2004_47_8_1783} to obtain this information from the scattered field  to gain insight into the composition of the scattered modes, and uncover the details of the physics.


\begin{acknowledgments}
This research was supported in part through the use of the Farber computer cluster and associated Information Technologies (IT) resources at the University of Delaware.
\end{acknowledgments}

\appendix

\section{Extension of the FD-PML Method to 3D}

The FD-PML method for phonons has previously been mathematically described and validated for one-dimension chains as well as in two-dimensions for the specific case of determining the scattering cross section of embedded nanoparticles with simple cubic lattice structure and second nearest neighbor interactions\cite{JAP_2015_118_9_094301}.  Here a more general mathematical description is given, valid for arbitrary 3D arrangements of atoms.  

The dynamic equations governing the displacements ($\tilde{u}$) of each atom in the simulation cell (for example, box 2 and 3 in Fig. \ref{fig:Computational_Domain}) can be written in the frequency domain for each atom under the harmonic approximation as

\begin{equation}
    (\mathbf{M}\omega^2+\mathbf{K})\tilde{\mathbf{u}}=0 
    \label{eqn:Motion_equation}
\end{equation}

where $\mathbf{M}$  is a diagonal mass matrix given by $\mathbf{M}=m_i\delta_{ij}$ (summation not implied) where $m_i$ is the mass associate with the $i^{th}$ degree of freedom (each atom has 3-degrees of freedom), and $\mathbf{K}$ are the interatomic force constants given by

\begin{equation}
    K_{ij}=\frac{\partial^2\Phi}{\partial u_i\partial u_j}
    \label{eqn:Spring_matrix}
\end{equation}
where  $\Phi$ is the interatomic potential and $u_i$ is the real-space displacement of the $i^{th}$ degree of freedom relative to the equilibrium position. These can be obtained from ab-initio calculation or based on some other description of the interatomic potentials; in the current paper, we used the simplest possible interaction, based on nearest-neighbor bonding via linear spring interactions.  The diagonal elements for $\mathbf{K}$  must obey $k_{ii}=-\sum_{i\ne j} k_{ij}$ . 

The frequency domain displacements, $\tilde{\mathbf{u}}$, are then divided into contributions from an incident wave (which is to be specified before the simulation) and a scattered wave (which is to be found as part of the solution procedure):  $\tilde{\mathbf{u}}=\tilde{\mathbf{u}}^{inc}+\tilde{\mathbf{u}}^{scat}$.     
Typically, the context for specifying the incident mode is self-evident, but the method provides tremendous flexibility as to how this is accomplished, and several reasonable variations are possible.  For example, if the objective of the computation is to simulate the transmission coefficient from one bulk material to another, the incident wave displacements would be specified as a plane wave obeying the dispersion relation in one of the materials, and would be set to zero everywhere the relation isn't expected to hold; so for a Si-Ge interdiffused interface, one could specify an incident wave within some portion of the purely Si region, and zero everywhere else (including the interdiffused zone).

As discussed thus far, the equations for the system would be given as
\begin{equation}
    (\mathbf{M}\omega^2+\mathbf{K})\tilde{u}^{scat}=-(\mathbf{M}\omega^2+\mathbf{K})\tilde{u}^{inc}
    \label{eqn:scattered_equation}
\end{equation}
However, the issue of how to handle the outer boundary conditions is also critical.  A typical incident plane wave carries energy into a system, and in most physical situations this energy is scattered or reflected by inhomogeneities.  This scattered energy must be allowed to escape the system.  For reflections from interfaces, the scattered modes are also expected to be infinitely extended in space, which is not possible to simulate numerically.  For scattered modes that do decay radially, the displacements are only zero infinitely far from the scattering center; numerically setting the scattered wave displacements to zero at any finite distance is equivalent to preventing energy from leaving the system.

To solve this issue, a special layer called a Perfectly Matched Layer (PML) is employed at the boundaries of the simulation cell to absorb the scattered wave. The total simulation domain can thus be decomposed into a Primary Domain (PD), which contains the scattering problem to be solved, and a PML domain (PMLD) forming boundary layers at the edges of the simulation domain.  A PML is designed to be impedance matched with the PD\cite{JCP_1994_114_2_0021,Chew1994}, and thus minimizes artificial reflections of the scattered wave back into the PD. PMLs were initially developed to aid solution of finite-difference time domain problems in electromagnetics, but have been applied to a variety of problems since.  Li et al\cite{PRB_2006_74_4_045418} have described their use in time-domain discrete atomic systems, and Kakodkar et al has extended this to the frequency domain \cite{JAP_2015_118_9_094301}.  Li et al \cite{PRB_2006_74_4_045418} has shown that for discrete atomic systems, the PMLD has nearly the same equations of motion as the original system, except that atoms in the PML possess an onsite potential and a damping force.  As such, for the FD-PML method, the full system of equations with the PML becomes

\begin{equation}
    (\mathbf{M}\omega^2+\mathbf{K}+\mathbf{M}\mathbf{\sigma}^2-2 i \mathbf{M}\mathbf{\sigma})\tilde{\mathbf{u}}^{scat}=\\-(\mathbf{M}\omega^2+\mathbf{K})\tilde{\mathbf{u}}^{inc}
    \label{eqn:PML_equation}
\end{equation}
 $\mathbf{\sigma}$  is a diagonal matrix of damping coefficients for each degree of freedom.  The value of $\mathbf{\sigma}$ can be chosen with considerable freedom, and need not be homogeneous.  We use an inhomogeneous $\mathbf{\sigma}$ , whose value for atoms inside the PMLD is given by

\begin{equation}
    \mathbf{\sigma}=\frac{\sigma_{max}|r_i|^2}{L_{PML}^2}\delta_{ij}
    \label{eqn:Damping_coefficient}
\end{equation}
where  $\sigma_{max}$ is the maximum value of the damping coefficient, $L_{PML}$  is the total length of the PML layer and the term $r_i$  represents the shortest distance of the $i^{th}$ atom from PD-PMLD boundary. For atoms inside the PD the damping coefficient is simply set to zero. Here we have chosen a parabolic distribution of the damping coefficient and a more comprehensive study for obtaining the PML parameters ($\sigma_{max},L_{PML}$)  is given in Kakodkar et al.\cite{JAP_2015_118_9_094301}. The following empirical relations were used throughout this manuscript to set the PML parameters and were found to result in artificial PML energy reflection coefficients of order $10^{-5}$.  

\begin{equation}
    \alpha_1= A\alpha_3+B,
    \label{eqn:alpha_1}
\end{equation}

\begin{equation}
    \alpha_2= 15(\cos^2\theta+0.5)
    \label{eqn:alpha_2}
\end{equation}

Where $\alpha_1, \alpha_2$ and $\alpha_3$ are dimensionless parameters given by $\alpha_1= k_z L_{PML}/(2\pi)$, $\alpha_2= (\sigma_{max} L_{PML})/ c_p$ and $\alpha_3= k_za/(2\pi)$. Here $k_z$ is the wavevector component in the direction normal to the interface, $c_p= \omega/||k||$ is the phase velocity and $\theta$ is the angle made by the incident wave-vector normal to the interface. The values of A and B used for this work are 27.8 and -1.3 respectively

Equation \ref{eqn:PML_equation} is a sparse, banded linear algebraic equation where the terms on the right hand side are known. The equation is solved numerically to obtain the scattered wave solution ($\tilde{u}_{scat}$).  Note that unlike the atomistic Green's function method, one does not need to store the inverted matrix which typically has order, $O(N^2)$, non-zero elements, where $N$ is the total number of atoms; for comparison, the initial matrix typically only has order, $O(N)$ non-zero elements.  Thus, iterative solvers enable very large system sizes to be simulated without a distributed memory architecture.

The total energy leaving the system through the scattered wave is then straightforwardly obtained. Since all the energy is absorbed by dampers in the PML, the total scattered energy, $Q$, is obtained inexpensively by summing the rate of energy absorption for each damper such that

\begin{equation}
    Q=\omega^2\sum_i {m_i \sigma_{ii} (\tilde{\mathbf{u}}^{scat}_i)^*}\tilde{\mathbf{u}}^{scat}_i
    \label{eqn:energy}
\end{equation}


\providecommand{\noopsort}[1]{}\providecommand{\singleletter}[1]{#1}%

\end{document}